\documentclass[amstex,twocolumn,showpacs,floats,floatfix,superscriptaddress,aps,pra]{revtex4}
\usepackage{amssymb}
\usepackage{amsmath}
\usepackage{calc}
\usepackage{graphicx}
\usepackage{bm}

\def\E{\mathcal{E}}
\def\const{\text{const}}
\def\e{\text{e}}
\def\m{\text{m}}
\def\r{\text{r}}
\def\w{\text{w}}

\begin{document}
\title{Mid-range adiabatic wireless energy transfer via a mediator coil}

\author{Andon A. Rangelov}
\affiliation{Department of Physics, Sofia University, 5 James
Bourchier blvd, 1164 Sofia, Bulgaria} \affiliation{Corresponding
author: rangelov@phys.uni-sofia.bg}

\author{Nikolay V. Vitanov}
\affiliation{Department of Physics, Sofia University, 5 James
Bourchier blvd, 1164 Sofia, Bulgaria}

\date{\today }

\begin{abstract}
A technique for efficient mid-range wireless energy transfer between
two coils via a mediator coil is proposed. By varying the coil
frequencies three resonances are created: emitter-mediator (EM),
mediator-receiver (MR) and emitter-receiver (ER). If the frequency
sweeps are adiabatic and such that the ER resonance precedes the MR
resonance, the energy flows sequentially along the chain
emitter-mediator-receiver. If the MR resonance precedes the ER
resonance, then the energy flows directly from the emitter to the
receiver via the ER resonance; then the losses from the mediator are
suppressed. This technique is robust to noise, resonant constraints
and external interferences.
\end{abstract}

\maketitle


Since the pioneering work of Tesla 
 the search for wireless power transfer is a hot topic for its vast potential to substitute electrical cables and batteries.
Until very recently traditional magnetic induction, wherein two
conductive coils induct each other, has been used to transfer energy
wirelessly over very short distances only. The coils do not make
direct electrical contact with each other but they must be very
close because
 the efficiency of power transfer drops by orders of magnitude when the distance between the coils becomes larger than their sizes.

Recently, some novel ideas for medium-range wireless energy transfer
have emerged \cite{Kurs,Karalis}. Specially designed magnetic
resonators achieve strong coupling between the coils that enable
efficient energy transfer over distances much larger than the size
of the coils \cite{Kurs,Karalis}. This technique, which demands the
same frequencies of the emitter and receiver coils, can be made more
robust to variations of the coil parameters by sweeping the emitter
frequency through resonance with the receiver frequency (or vice
versa) \cite{Rangelov} in a fashion reminiscent of adiabatic passage
through a level crossing in quantum physics \cite{LZSM}.

A recent theoretical paper \cite{Hamam} proposed a setup of two
identical coils (emitter and receiver)
 strongly coupled to an intermediate coil (mediator) of different properties but with the same intrinsic frequency.
The energy transfer occurs due to adiabatic following of an
instantaneous (``dark'') eigenstate of the three-coil system,
 an idea similar to the quantum three-state technique of stimulated Raman adiabatic passage (STIRAP) \cite{Gaubatz,VitanovA,VitanovB}.
In this technique both the emitter and the receiver should be
rotated in synchro in order to engineer time-dependent couplings
that reduce the energy stored at the mediator \cite{Hamam}. It is,
however, crucial to maintain an exact resonance between the emitter
and receiver frequencies \cite{Hamam};
 a frequency mismatch, e.g., due to differences the coils or random noise (which may be caused by external objects near the coils), reduces the transfer efficiency significantly.

In this paper, we propose a different approach to wireless energy
transfer between two coils via a mediator coil
 by using the ideas of adiabatic population transfer in a three-state quantum system with crossing energies \cite{Broers,Unanyan,Ivanov}.
This technique promises to be both efficient and robust against
variations of the parameters, such as the intrinsic frequencies of
the coils and the couplings between them.


\begin{figure}[t]
\centerline{\includegraphics[width=0.65\columnwidth]{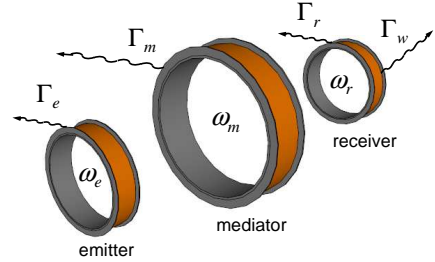}}
\caption{(Color online) Three coils with intrinsic frequencies
$\omega_{\e}$, $\omega_{\m}$ and $\omega_{\r}$. The large size of
the mediator leads to strong couplings with the others coils, but
also to larger loss rate $\Gamma_{\m}$
 than the loss rates $\Gamma_{\e}$ and $\Gamma_{\r}$ of the emitter and receiver.}
\label{Fig1}
\end{figure}

We consider three parallel helix coils as shown in Fig.~\ref{Fig1}:
two small coils --- an emitter and a receiver --- and a large
mediator, which induces strong couplings.
We follow the description of the coupled mode theory 
 as presented by Hamam \emph{et al.} \cite{Hamam}.
The energy transfer, in the strong-coupling regime, is described by
a set of three coupled differential equations (in matrix form)
\cite{Hamam},
\begin{subequations}\label{three states system}
\begin{equation}
i\frac{d\mathbf{A}(t)}{dt}=\mathbf{H}(t)\mathbf{A}(t),
\end{equation}
where $\mathbf{A}=\left[ a_{\e}(t),a_{\m}(t),a_{\r}(t)\right]^T$ and
\begin{equation}
\mathbf{H}=\left[\begin{array}{ccc}
\omega_{\e}-i\Gamma_{\e} & \kappa_{\e\m} & 0 \\
\kappa_{\e\m} & \omega_{\m}-i\Gamma_{\m} & \kappa_{\m\r} \\
0 & \kappa_{\m\r} & \omega_{\r}-i\Gamma_{\r}-i\Gamma_{\w}
\end{array}\right].
\end{equation}
\end{subequations}
Here $a_{\e}(t)$, $a_{\m}(t)$ and $a_{\r}(t)$ are defined so that
the energies contained in the emitter, the mediator and the receiver
are, respectively, $|a_{\e}(t)|^{2}$, $|a_{\m}(t)|^{2}$ and
$|a_{\r}(t)|^{2}$,
 while $\Gamma_{\e}$, $\Gamma_{\m}$\ and $\Gamma_{\r}$ are the corresponding intrinsic loss rates (due to absorption and radiation), 
 and $\omega_{\e}(t)$, $\omega_{\m}(t)$ and $\omega_{\r}(t)$ are the corresponding intrinsic frequencies.
The extraction of work from the receiver is described by the term
$\Gamma_{\w}$. The coupling coefficient between the emitter and the
mediator is $\kappa_{\e\m}$ and the one between the mediator and the
receiver is $\kappa_{\m\r}$. The null elements in the matrix
$\mathbf{H}(t)$ derive from the assumption that the mediator is much
larger than the emitter and the receiver;
 hence the direct emitter-receiver coupling is negligible compared to the couplings $\kappa_{\e\m}$ and $\kappa_{\m\r}$ and is neglected.

The efficiency of the energy transfer is described by the efficiency
coefficient $\eta$ \cite{Kurs,Karalis,Rangelov},
 which is the ratio between the work extracted from the receiver for the time interval $t_{i}-t$ divided by the total energy (absorbed and radiated) for the same time interval,
\begin{equation}
\eta =\frac{\Gamma_{\w} \E_\r(t)}
{\Gamma_{\e}\E_\e(t)+\Gamma_{\m}\E_\m(t)+(\Gamma_{\r}+\Gamma_{\w})\E_\r(t)},
\label{efficiency coefficient}
\end{equation}
where $\E_k(t) = \int_{t_{i}}^{t}|a_{k}(t)|^{2}\text{d}t$
($k=\e,\m,\r$).

Equation \eqref{three states system} is identical to the time-dependent Schr\"{o}dinger equation for a three-state quantum system 
 which is studied in great detail \cite{VitanovA,VitanovB}; the vector $\mathbf{A}(t)$ and the driving matrix $\mathbf{H}(t)$ correspond to the quantum state vector and the Hamiltonian, respectively.
By definition, in the adiabatic limit the system stays in an
eigenvector of $\mathbf{H}(t)$. We first assume that the loss rates
$\Gamma_{\e}$, $\Gamma_{\m}$, $\Gamma_{\r}$ and $\Gamma_{\w}$ are
all zero;
 then the quantity $|\mathbf{A}(t)|^{2}=|a_{\e}(t)|^{2}+|a_{\m}(t)|^{2}+|a_{\r}(t)|^{2}$ is conserved, like the total population in a coherently driven quantum system.

The key of our proposal is the assumption that the frequency of the
mediator coil is fixed, while the frequencies of the emitter and
receiver coils change in time in opposite directions,
\begin{subequations}\label{sweep}
\begin{align}
\omega_\e(t) &= \omega_\m + \delta + \alpha^{2}t, \\
\omega_\m &= \const,\\
\omega_\r(t) &= \omega_\m + \delta - \alpha^{2}t,
\end{align}
\end{subequations}
where $\delta$ is a suitably chosen static frequency offset, which
controls the energy flow. For the sake of generality, we take
hereafter $\alpha>0$ as the unit of frequency and $1/\alpha $ as the
unit of time. We assume that the fixed mediator frequency
$\omega_{\m}$ is much larger than  the terms $\delta\pm\alpha^2 t$
in $\omega_{\e}(t)$ and $\omega_{\r}(t)$;
 therefore the couplings $\kappa_{\e\m}$ and $\kappa_{\m\r}$ can be assumed constant for the sake of simplicity.
Taking the variations of the couplings due to the frequency sweeps
into account does not change the results significantly.

\begin{figure}[t]
\centerline{\includegraphics[width=0.9\columnwidth]{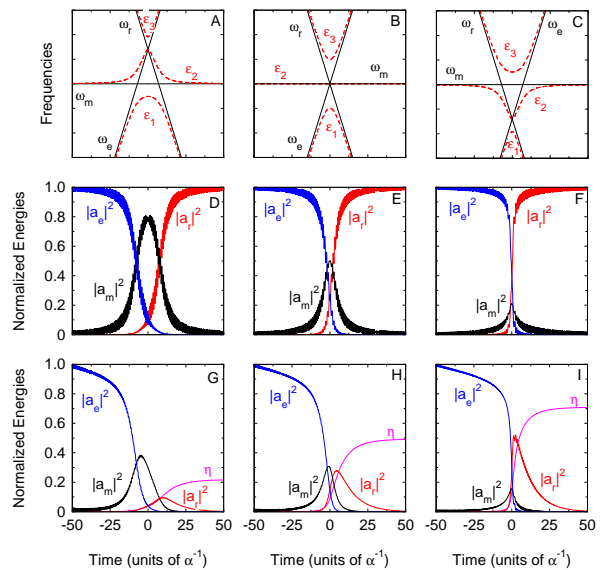}}
\caption{(Color online) Wireless energy transfer. Left column:
sequential transfer ($\delta=7\alpha$); middle column: bow-tie
transfer ($\delta=0$); right column: direct transfer
($\delta=-7\alpha$). Top frames: Diagonal elements (solid lines) and
eigenvalues (dashed lines) of $\mathbf{H}$ of Eq.~\eqref{three
states system}. Middle frames: The energies in the coils calculated
numerically from Eq.~\eqref{three states system} with no losses,
$\Gamma_{\e}=\Gamma_{\m}=\Gamma_{\r}=\Gamma_{\w}=0$
 for $\kappa_{\e\m}=\kappa_{\m\r}=3.5\alpha$.
Bottom frames: The energies in the coils and the efficiency
coefficient $\eta$ for $\Gamma_{\e}=\Gamma_{\r}=0.001\alpha $,
$\Gamma_{\m}=\Gamma_{\w}=0.05\alpha $. } \label{Fig2}
\end{figure}

For $\delta\neq 0$, the intrinsic frequencies of the three coils cross each other at three different instants of times, 
 thereby creating a triangle crossing pattern \cite{Broers,Unanyan,Ivanov}, shown in Fig.~\ref{Fig2} (top frames).
For $\delta=0$, the three frequencies cross at the same instant of
time, thereby creating a bow-tie crossing (frame B) \cite{bow-tie}.
These crossing patterns allow us to design recipes for efficient
adiabatic wireless energy transfer,
 in analogy to adiabatic passage techniques in quantum physics \cite{VitanovA,VitanovB,Broers,Unanyan,Ivanov}.

The proposed technique is illustrated in Fig.~\ref{Fig2}. The top
frames show the evolution of the intrinsic frequencies of the three
coils $\omega_k(t)$  ($k=\e,\m,\r$) and the eigenvalues
$\varepsilon_n(t)$  ($n=1,2,3$)  (the eigenfrequencies) of
$\mathbf{H}(t)$
 of Eq.~\eqref{three states system}.
The three columns of frames differ by the offset frequency $\delta$:
$\delta>0$ in the left column, $\delta=0$ in the middle column and
$\delta<0$ in the right column. In the beginning and the end each
eigenfrequency $\varepsilon_n(t)$ coincides with one of the
intrinsic frequencies $\omega_k(t)$ of the coils,
 while in between each $\varepsilon_n(t)$ is a superposition of $\omega_\e(t)$, $\omega_\m$ and $\omega_\r(t)$.
In the adiabatic limit, the system follows an eigenstate of
$\mathbf{H}(t)$ and the eigenfrequency of the system at any instant
of time
 is equal to the (time-dependent) eigenfrequency in which it is initially, i.e. $\varepsilon_1(t)$.
However, the composition of $\varepsilon_1(t)$ is different for
different $\delta$ because the topology of the crossings differ.

For $\delta>0$ (left column of frames), the frequency resonances
occur in the intuitive manner: first the emitter-mediator resonance
and then the mediator-receiver resonance. The eigenfrequency
$\varepsilon_1(t)$ is initially nearly equal to $\omega_\e$, at
intermediate times to $\omega_\m$ and in the end to $\omega_\r$.
Consequently, the energy flows sequentially from the emitter to the
mediator and then to the receiver coils; this sequential flow is
demonstrated in frame D. This two-step scheme extends the
single-step adiabatic wireless energy transfer proposed earlier
\cite{Rangelov}.

For $\delta=0$ (middle column of frames) --- the bow-tie crossing
--- all resonances occur at the same time. The composition of
$\varepsilon_1(t)$ is similar as for $\delta>0$, the difference
being the lesser contribution of the mediator frequency $\omega_\m$
at intermediate times. Consequently, the mediator receives less
transient energy, as seen in frame E.

For $\delta<0$ (right column of frames), the frequency resonances
occur in a counterintuitive manner, with the mediator-receiver
resonance occurring before the emitter-mediator resonance. In fact,
the system never comes close to the ``nearest-neighbor''
emitter-mediator and mediator-receiver resonances but it passes
through the direct emitter-receiver resonance (frame C). The
eigenfrequency $\varepsilon_1(t)$ remains very far from the mediator
frequency $\omega_\m$;
 consequently, very little energy attends transiently the mediator coil, as seen in frame F.
This feature is reminiscent of the energy transfer in the
STIRAP-based technique \cite{Hamam}.

The coil parameters in Fig.~\ref{Fig2} --- the intrinsic
frequencies, the couplings and the sweep rate --- are chosen in a
such way that in the absence of losses the evolution is adiabatic
 and hence the energy is transferred almost completely to the receiver coil in the end, as seen in frames D, E and F of Fig.~\ref{Fig2}.
However, the temporary transfer of energy to the mediator coil makes
the energy transfer efficiency very different in the presence of
losses, as seen in the bottom frames of Fig.~\ref{Fig2}. For
$\delta>0$, a large amount of energy visits the mediator where is
subjected to strong dissipation and most of it is lost before it has
the chance to reach the receiver (frame G). For $\delta=0$, less
energy attends the mediator coil and hence some more energy reaches
the receiver (frame H). For $\delta<0$, only little energy visits
the mediator coil and a significant amount of energy is transferred
to the receiver (frame I).

We therefore identify the direct adiabatic path, occurring for
$\delta<0$, as the optimal path for energy transfer from the emitter
to the receiver.

Next we turn our attention to the conditions for adiabatic evolution
in the three distinct cases $\delta>0$, $\delta=0$ and $\delta<0$.
By using the formula for the transition probability for the
Landau-Zener-St\"uckelberg-Majorana (LZSM) model \cite{LZSM},
$p=1-\exp(-2\pi\kappa^2/\alpha^2)$,
 we find that the condition for transition probability larger than $1-\epsilon$ at each crossing reads
\begin{equation}\label{adiabatic conditions}
\frac{\kappa}{\alpha} > \sqrt{\frac{\ln (1/\epsilon)}{2\pi}},
\end{equation}
where $\kappa=\kappa_{\e\m}$ for the emitter-mediator resonance and
$\kappa=\kappa_{\m\r}$ for the mediator-receiver resonance. For the
bow-tie crossing, the condition is (for
$\kappa_{\e\m}=\kappa_{\m\r}$) \cite{bow-tie}
\begin{equation}\label{adiabatic conditions bow-tie}
\frac{\kappa}{\alpha} > \sqrt{\frac{\ln (1/\epsilon)}{\pi}}.
\end{equation}
For the direct emitter-receiver resonance, we can use the LZSM
formula again, by taking into account that the sweep rate is
$2\alpha$ and the effective emitter-receiver coupling is
 $\kappa=\sqrt{\kappa_{\e\m}^2+\kappa_{\m\r}^2+\delta^2/4}-\delta/2$ \cite{Ivanov},
\begin{equation}\label{adiabatic condition E-R}
\frac{\kappa}{\alpha} > \sqrt{\frac{2\ln (1/\epsilon)}{\pi}}.
\end{equation}
In addition to these conditions, efficient energy transfer requires
that the loss rates be small compared to the couplings and the
interaction time $T$,
\begin{equation}\label{adiabatic condition Gamma}
\Gamma_k \ll 1/T,\kappa.
\end{equation}
One can readily verify that all these conditions are satisfied for
the parameters in Fig.~\ref{Fig2}.

In conclusion, the proposed technique for adiabatic wireless energy
transfer between an emitter and a receiver coil via a larger
mediator coil allows one to transfer energy with high efficiency by
varying the intrinsic frequencies of the coils in the
counterintuitive way, in which the mediator-receiver resonance
occurs before the emitter-mediator resonance. Then only a small
amount of energy visits the mediator and hence the losses from the
mediator are minimized. The presence of a large mediator coil allows
one to increase the coil-coil couplings and therefore to increase
the distance between the emitter and the receiver compared to the
simple two-coil emitter-receiver setup \cite{Rangelov}. Compared to
the STIRAP-based energy transfer technique \cite{Hamam}, the present
technique does not require identical coil frequencies and
time-varying couplings, and therefore it may be easier to implement.

This work is supported by the European Commission project FASTQUAST,
and the Bulgarian National Science Fund grants D002-90/08 and
DMU-03/103.



\begin{thebibliography}{99}

\bibitem{Kurs} A. Kurs, A. Karalis, R. Moffatt, J. D. Joannopoulos, P. Fisher, and M. Solja\v{c}i\'{c}, Science \textbf{317}, 83 (2007).

\bibitem{Karalis} A. Karalis, J. D. Joannopoulos, and M. Solja\v{c}i\'{c}, Ann. Phys. \textbf{323}, 34 (2008).

\bibitem{Rangelov} A. A. Rangelov, H. Suchowski, Y. Silberberg, and N. V. Vitanov, Ann. Phys. \textbf{326}, 626 (2011).

\bibitem{LZSM} L. D. Landau, Physik Z. Sowjetunion \textbf{2}, 46 (1932); C. Zener, Proc. R. Soc. Lond. Ser. A \textbf{137}, 696 (1932); E. C. G. St\"{u}ckelberg, Helv. Phys. Acta \textbf{5}, 369 (1932); E. Majorana, Nuovo Cimento \textbf{9}, 43 (1932).

\bibitem{Hamam} R. E. Hamam, A. Karalis, J. D. Joannopoulos, and M. Solja\v{c}i\'{c}, Ann. Phys. \textbf{324}, 1783 (2009).

\bibitem{Gaubatz} U. Gaubatz, P. Rudecki, S. Schiemann, and K. Bergmann, J. Chem. Phys. \textbf{92}, 5363 (1990).


\bibitem{VitanovA} N. V. Vitanov, T. Halfmann, B. W. Shore, and K. Bergmann, Annu. Rev. Phys. Chem. \textbf{52}, 763 (2001).

\bibitem{VitanovB} N. V. Vitanov, M. Fleischhauer, B. W. Shore, and K. Bergmann, Adv. At. Mol. Opt. Phys. \textbf{46}, 55 (2001).

\bibitem{Broers} B. Broers, L. D. Noordam, and H. B. van Linden van den Heuvell, Phys. Rev. A \textbf{46}, 2749 (1992).

\bibitem{Unanyan} R. G. Unanyan, N.V. Vitanov, and K. Bergmann, Phys. Rev. Lett. \textbf{87}, 137902 (2001).

\bibitem{Ivanov} S. S. Ivanov and N. V. Vitanov, Phys. Rev. A \textbf{77}, 023406 (2008).

\bibitem{bow-tie} C. E. Carroll and F. T. Hioe, J. Phys. A: Math. Gen. \textbf{19}, 1151 (1986);
 C. E. Carroll and F. T. Hioe, J. Phys. A: Math. Gen. \textbf{19}, 2061 (1986);
 V. N. Ostrovsky and H. Nakamura, J. Phys. A \textbf{30}, 6939 (1997). 

\end{thebibliography}
\end{document}